\documentclass[english]{IEEEtran}
\usepackage[T1]{fontenc}
\usepackage[latin9]{inputenc}
\usepackage{color}
\usepackage{float}
\usepackage{units}
\usepackage{amstext}
\usepackage{graphicx}
\usepackage{esint}

\makeatletter
\usepackage{hyperref}

\makeatother

\usepackage{babel}
\begin{document}

\title{Identifying Useful Statistical Indicators of Proximity to Instability
in Stochastic Power Systems}

\author{Goodarz Ghanavati, \emph{Student Member, IEEE}, Paul D.~H.~Hines,
\emph{Senior Member, IEEE}, Taras I.~Lakoba%
\thanks{This work was supported by the US DOE, award \#DE-OE0000447, and by
the US NSF, award \#ECCS-1254549.

G.~Ghanavati, P.~D.~H.~Hines, T.~I.~Lakoba are with the College
of Engineering and Mathematical Sciences, University of Vermont, Burlington,
VT (e-mail: gghanava@uvm.edu; paul.hines@uvm.edu; tlakoba@uvm.edu).%
}}
\maketitle
\begin{abstract}
Prior research has shown that autocorrelation and variance in voltage
measurements tend to increase as power systems approach instability.
This paper seeks to identify the conditions under which these statistical
indicators provide reliable early warning of instability in power
systems. First, the paper derives and validates a semi-analytical
method for quickly calculating the expected variance and autocorrelation
of all voltages and currents in an arbitrary power system model. Building
on this approach, the paper describes the conditions under which filtering
can be used to detect these signs in the presence of measurement noise.
Finally, several experiments show which types of measurements are
good indicators of proximity to instability for particular types of
state changes. For example, increased variance in voltages can reliably
indicate the location of increased stress, while growth of autocorrelation
in certain line currents is a reliable indicator of system-wide instability.\end{abstract}

\begin{IEEEkeywords}
Power system stability, phasor measurement units, time series analysis,
stochastic processes, principal component analysis, autocorrelation,
critical slowing down.
\end{IEEEkeywords}

\section{Introduction}

To make optimal use of constrained infrastructure, power systems frequently
operate near their stability limits. Bifurcation theory provides a
framework for understanding these instabilities \cite{alvarado1994computation}\nocite{avalos2009equivalency,grijalva2012individual}--\cite{perninge2010risk}
and has motivated the development of new methods for online stability
monitoring \cite{haque2003line}\nocite{Browne2008}--\cite{kamwa2011robust}.

This existing work has largely focused around deterministic power
system models. However, real power systems are constantly influenced
by stochastic perturbations in load and (increasingly) variable renewable
generation. Because random fluctuations can substantially change the
stability properties of a system \cite{wangfokker}, several have
proposed the use of stochastic approaches to stability analysis~(e.g.,
\cite{nwankpa1992stochastic}\nocite{Anderson:1983}\nocite{haesen2009probabilistic,perninge2010risk,bu2012probabilistic,munoz2013affine}--\cite{huang2013quasi}). 

Indeed, outside of the power systems literature, there is growing
evidence that complex systems show statistical early warning signs
as they approach instability~\cite{scheffer2009early,Scheffer:2012}.
This phenomenon, known as critical slowing down (CSD)~\cite{mori1963relaxation},
is the tendency of a dynamical system to return to equilibrium more
slowly in response to perturbations as it approaches a critical bifurcation.
Increasing autocorrelation and variance in measurements, two common
signs of \textcolor{black}{CSD, have been shown to }signal proximity
to critical transitions in a variety of dynamical systems \cite{scheffer2009early}.
However, not all measurements show these signs early enough to provide
warning with sufficient time to take mitigating actions~\cite{boerlijst2013catastrophic}.
Understanding which variables provide useful early warning of instability
is necessary for the practical application of these concepts. Doing
so requires a detailed knowledge of how autocorrelation and variance
change as a system's state changes.

A few papers have studied the properties of variance and autocorrelation
as indicators of instability in power systems. Reference~\cite{Cotilla2012}
showed, using simulations, that variance and autocorrelation of bus
voltages increase before bifurcation. Reference~\cite{podolsky2013}
derives the autocorrelation function of a power system's state vector
near a saddle-node bifurcation and uses the result to estimate the
collapse probability for power systems. In~\cite{Dhople2013}, a
framework is proposed to study the impact of stochastic power injections
on power system dynamics by computing the moments of the states. In~\cite{ghanavati2013understanding},
the authors showed that for some state variables, increases in autocorrelation
and variance appear only when a power system is very close to the
\textcolor{black}{bifurcation,} indicating that CSD does not always
provide useful early warning of instability. Reference~\cite{yuan2014stochastic}
calculates the variance of state variables to analyze the impact of
wind turbine mechanical power input fluctuations on small-signal stability. 

The goal of this paper is to present a general method for estimating
the autocorrelation and variance of state variables from a power system
model and to use the results to determine which variables in a power
system provide useful early warning of critical transitions in the
presence of measurement noise. To this end, Sec.~\ref{sec:Analytical}
presents a semi-analytical method for calculating the variance and
autocorrelation of algebraic and differential variables. This method
enables the fast calculation of voltage and current statistics for
many potential operating scenarios in large power systems, and unlike
the method in \cite{podolsky2013}, is not limited to the immediate
vicinity of a bifurcation. Sec.~\ref{sec:Useful-early-signs} illustrates
the method using the 39-bus test case and shows that some variables
are better indicators of proximity to instability than others. Sec.~\ref{sec:Detectability}
extends the analysis to systems with measurement noise and presents
a method for detecting CSD in the presence of measurement noise. Sec.~\ref{sec:Stressed-Area}
uses this approach to identify stressed areas in a power network.
Finally, our conclusions are presented in Sec.~\ref{sec:Conclusions}.

\section{Calculation of Autocorrelation and Variance in Multimachine Power
Systems \label{sec:Analytical}}

This section presents a semi-analytical method for the fast calculation
of variance $\left(\sigma^{2}\right)$ and autocorrelation $\left(R\left(\Delta t\right)\right)$
of bus voltage magnitudes and line currents in power system. Fluctuations
of load and generation are well known sources of stochasticity in
power systems. While this section models only randomness in load,
the method can be readily extended to the case of stochasticity in
power injections.

\subsection{System Model\label{sub:System-Model}}

Adding stochastic load to the set of general differential-algebraic
equations (DAE) that model a power system gives: 
\begin{eqnarray}
\dot{\underline{x}} & = & f\left(\underline{x},\underline{y}\right)\label{eq:diff}\\
0 & = & g\left(\underline{x},\underline{y},\underline{u}\right)\label{eq:alg}
\end{eqnarray}
where $f,g$ represent differential and algebraic equations, $\underline{x},\underline{y}$
are vectors of differential and algebraic variables (generator rotor
angles, bus voltage magnitudes, etc.), and $\underline{u}$ is the
vector of load fluctuations. The algebraic equations consist of nodal
power flow equations and static equations for components such as generator,
exciter, and turbine governor. The differential equations describe
the dynamic behavior of the equipment. In this paper, for modeling
load fluctuations, we take an approach similar to \cite{perninge2010risk},
\cite{hauer2007} and assume that load fluctuations $\underline{u}$
follow the Ornstein--Uhlenbeck process:
\begin{equation}
\dot{\underline{u}}=-E\underline{u}+\underline{\xi}\label{eq:load_corr}
\end{equation}
where $E$ is a diagonal matrix whose diagonal entries equal $t_{\textnormal{{corr}}}^{-1}$,
where $t_{\textnormal{{corr}}}$ is the correlation time of the load
fluctuations, and $\underline{{\xi}}$ is a vector of independent
Gaussian random variables:
\begin{eqnarray}
\textnormal{{E}}\left[\underline{\xi}\left(t\right)\right] & = & 0\label{eq:xi1}\\
\textnormal{\textnormal{{E}}}\left[\xi_{i}\left(t\right)\xi_{j}\left(s\right)\right] & = & \delta_{ij}\sigma_{\xi}^{2}\delta_{I}(t-s)\label{eq:xi2}
\end{eqnarray}
where $t,s$ are two arbitrary times, $\delta_{ij}$ is the Kronecker
delta function, $\sigma_{\xi}^{2}$ is the intensity of noise and
$\delta_{I}$ represents the unit impulse (delta) function. Equations
(\ref{eq:diff})--(\ref{eq:load_corr}) form the set of SDAEs that
models a power system with stochastic load.

We also consider the frequency-dependence of loads, which can measurably
impact the statistics of voltage magnitudes~\cite{ghanavati2013understanding}.
Loads are thus modeled as follows \cite{berg1973power,Milano2008}:
\begin{eqnarray}
\Delta\omega & = & \frac{1}{2\pi f_{n}}\frac{d\left(\theta-\theta^{0}\right)}{dt}\label{eq:freq1}\\
P & = & P^{0}\left(1+\Delta\omega\right)^{\beta_{P}}\label{eq:freq2}\\
Q & = & Q^{0}\left(1+\Delta\omega\right)^{\beta_{Q}}\label{eq:freq3}
\end{eqnarray}
where $\Delta\omega$ is the frequency deviation at the load bus,
$\theta^{0},P^{0},Q^{0}$ are the baseline voltage angle, active and
reactive power of each load, $\beta_{P},\beta_{Q}$ are exponents
that determine the level of frequency dependence, $f_{n}$ is the
nominal frequency and $\theta$ is the bus voltage angle.

Using this model, we studied the New England 39-bus test case \cite{pai1989energy}.
As load increases,  a Hopf bifurcation occurs just before the nose
of the PV curve (see \cite{lerm2003multiparameter,rosehart1999bifurcation}).

\subsection{Solution Method\label{sub:Solution-Method}}

Linearizing (\ref{eq:alg}) gives the following:
\begin{equation}
\Delta\underline{y}=\left[\begin{array}{cc}
-g_{y}^{-1}g_{x} & -g_{y}^{-1}g_{u}\end{array}\right]\left[\begin{array}{c}
\Delta\underline{x}\\
\Delta\underline{u}
\end{array}\right]\label{eq:Y-XU}
\end{equation}
where $g_{x},g_{y},g_{u}$ are the Jacobian matrices of $g$ with
respect to $\underline{x},\underline{y},\underline{u}$. Linearizing
(\ref{eq:diff}) and (\ref{eq:load_corr}) and eliminating $\Delta\underline{y}$
via (\ref{eq:Y-XU}) gives the following:
\begin{eqnarray}
\left[\begin{array}{c}
\Delta\underline{\dot{x}}\\
\Delta\underline{\dot{u}}
\end{array}\right] & = & \left[\begin{array}{cc}
A_{s} & -f_{y}g_{y}^{-1}g_{u}\\
0 & -E
\end{array}\right]\left[\begin{array}{c}
\Delta\underline{x}\\
\Delta\underline{u}
\end{array}\right]+\label{eq:SDE-linear}\\
 &  & \left[\begin{array}{c}
0\\
\textnormal{{\ensuremath{I_{n}}}}
\end{array}\right]\underline{\xi}\nonumber 
\end{eqnarray}
where $f_{x},f_{y}$ are the Jacobian matrices of $f$ with respect
to $\underline{x},\underline{y}$ and $A_{s}=f_{x}-f_{y}g_{y}^{-1}g_{x}$;
$\textnormal{{\ensuremath{I_{n}}}}$ is an identity matrix, with $n$
being the length of $\underline{u}$. If we let $\underline{z}=\left[\begin{array}{cc}
\Delta\underline{x} & \Delta\underline{u}\end{array}\right]^{T}$, (\ref{eq:SDE-linear}) can be re-written in the standard form:
\begin{equation}
\underline{\dot{z}}=A\underline{z}+B\underline{\xi}\label{eq:Ornsterin-Uhl}
\end{equation}

The covariance matrix of $\underline{z}$ ($\sigma_{\underline{z}}$)
satisfies the Lyapunov equation \cite{gardiner2012handbook}:
\begin{equation}
A\sigma_{\underline{z}}+\sigma_{\underline{z}}A^{T}=-BB^{T}\label{eq:cov_mat}
\end{equation}
which can be solved efficiently in $O\left(n^{3}\right)$ operations
using MATLAB's \verb+lyap+ function. To stress the difference between
the solution from (\ref{eq:cov_mat}) and the results of direct numerical
simulation of (\ref{eq:diff})--(\ref{eq:load_corr}), we will refer
to the former solution as semi-analytical.

The stationary autocorrelation matrix can be computed given $\sigma_{\underline{z}}$
and an equation from \cite{gardiner2012handbook}:
\begin{equation}
\textnormal{{E}}\left[\underline{z}\left(t\right)\underline{z}^{T}\left(s\right)\right]=\exp\left[-A\left|\Delta t\right|\right]\sigma_{\underline{z}}\label{eq:corr_z}
\end{equation}
where $\Delta t=t-s$. From (\ref{eq:cov_mat})~and~(\ref{eq:corr_z})
the normalized autocorrelation function of $z_{i}$ can be calculated:
\begin{equation}
R_{z_{i}}\left(\Delta t\right)=\textnormal{E}\left[z_{i}\left(t\right)z_{i}^{T}\left(s\right)\right]/\sigma_{z_{i}}^{2}\label{eq:norm_corr_z}
\end{equation}
The covariance matrix of the algebraic variables, $\sigma_{\underline{\Delta y}}$,
is found from (\ref{eq:Y-XU}) and (\ref{eq:cov_mat}):
\begin{equation}
\sigma_{\Delta\underline{{y}}}=K\sigma_{\underline{{z}}}K^{T}\label{eq:cov_y}
\end{equation}
where $K$ is the matrix from (\ref{eq:Y-XU}). Similarly, the autocorrelation
function of $\Delta\underline{y}(t)$ is:
\begin{equation}
\textnormal{E}\left[\Delta\underline{y}\left(t\right)\Delta\underline{y}^{T}\left(s\right)\right]=K\cdot\textnormal{{E}}\left[\underline{z}\left(t\right)\underline{z}^{T}\left(s\right)\right]K^{T}\label{eq:corr_y}
\end{equation}
Finally, the covariance and autocorrelation matrices for voltage magnitudes
are a subset of the matrices from (\ref{eq:cov_y}) and~(\ref{eq:corr_y}).

Fluctuation-induced deviations of the current magnitudes, $\Delta I_{ik}$,
in a line between buses $i$ and $k$ can be found by linearizing
the following:
\begin{equation}
I_{ik}=Y_{ii}V_{i}e^{j\left(\phi_{ik}-\phi_{ik}+\theta_{i}-\theta_{k}\right)}+Y_{ik}V_{k}\label{eq:line_cur}
\end{equation}
where $I_{ik}$ is the magnitude of the current of the line between
buses $i,k$; $V_{i},\theta_{i}$ are the voltage magnitude and angle
of bus~$i$; $Y_{ii},\phi_{ii}$ and $Y_{ik},\phi_{ik}$ are magnitudes
and angles of the diagonal and off-diagonal elements of the $Y_{BUS}$
matrix. By linearizing (\ref{eq:line_cur}) one can find $\Delta\underline{I}$
from $\Delta y$ and then compute the covariance and autocorrelation
matrices of $\Delta\underline{I}$ from equations similar to (\ref{eq:cov_y})
and (\ref{eq:corr_y}).

Comparing the semi-analytical method with the numerical solution shows
that the former is significantly more time-efficient. For the numerical
simulations in this paper, we solved (\ref{eq:diff})--(\ref{eq:load_corr})
using the trapezoidal DAE solver in the Power System Analysis Toolbox
(PSAT) \cite{milano2005open}. To find numerical values for $\sigma^{2}$
and $R(\Delta t)$ we ran 100 240s simulations, with an integration
step size of 0.01s, and then computed the statistics. For the 39-bus
case with 140 variables, solving for $\sigma_{z}^{2}$ using the semi-analytical
method took approximately 0.08s, whereas calculating the variances
using numerical simulations took about 24 hours.

\section{Useful early warning signs: voltage magnitudes and line currents\label{sec:Useful-early-signs}}

This section applies the method in Sec.~\ref{sec:Analytical} to
calculate the autocorrelation and variance of voltages and currents
in the 39-bus test case. These results are subsequently used to identify
particular locations and variables in which the statistical early-warning
signs are most clearly observable.

\subsection{Autocorrelation and Variance of Voltages \label{sub:Autoco-Var-Voltages}}

Using the methods described in Sec.~\ref{sec:Analytical}, we calculated
$\sigma^{2},\, R\left(\Delta t\right)$ of bus voltage magnitudes
in the 39-bus test case both semi-analytically and numerically using
PSAT. In order to see how these statistics change as the system state
moves toward the bifurcation, we increased all loads uniformly, multiplying
each load by the same factor. For the correlation time and intensity
of noise we used: $t_{corr}=1\textnormal{{s}}$ and $\sigma_{u}^{2}=10^{-4}$~pu.
The values of $\beta_{P},\beta_{Q}$ in (\ref{eq:freq2}), (\ref{eq:freq3})
were chosen randomly from within $\left[2,3\right]$ and $\left[1,2\right]$,
respectively \cite{berg1973power}. For all results in this paper,
we chose the autocorrelation time lag $\Delta t=0.2\textnormal{{s}}$\textcolor{black}{,
based on the criteria for choosing an optimal $\Delta t$ in \cite{ghanavati2013understanding}.}

\begin{figure}[H]
\begin{centering}
\includegraphics[width=1\columnwidth,height=0.2\textheight]{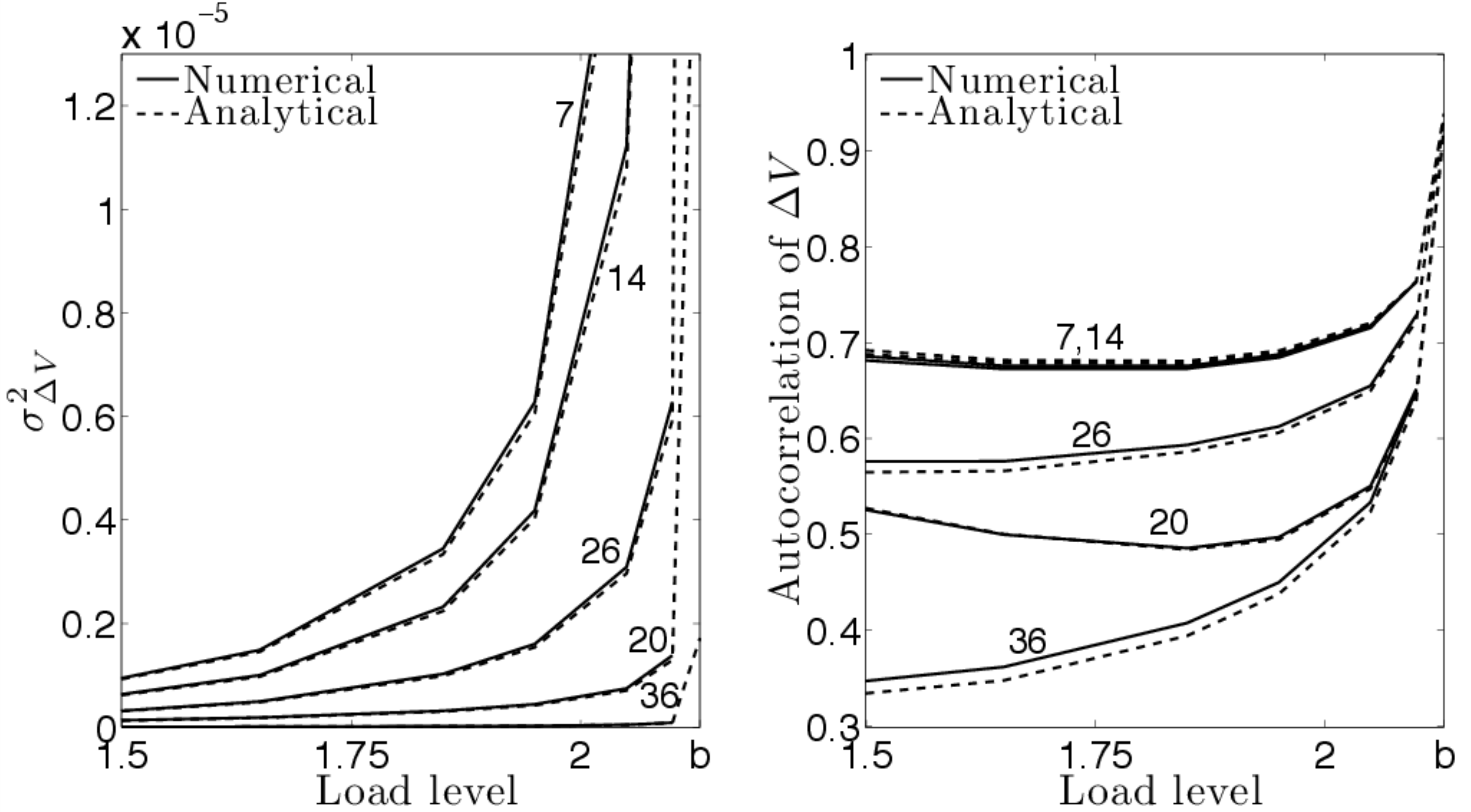}
\par\end{centering}

\vspace{-.1in}\protect\caption{\label{fig:arvar_vmag_39bus}Variance and autocorrelation of voltage
magnitudes for five buses in the 39-bus test case versus load level.
Load level is the ratio of the system loads to their nominal values.
b denotes the bifurcation point. The bus number associated with each
curve is shown next to it. Here and everywhere below the autocorrelation
time lag $t-s=0.2\textnormal{{s}}$.}

\vspace{-.1in}
\end{figure}

Fig.~\ref{fig:arvar_vmag_39bus} shows several typical, illustrative
examples of how $\sigma^{2},\, R\left(\Delta t\right)$ of bus voltage
magnitudes depend on load level in the 39-bus case.  These results
 show that, as anticipated from CSD theory, both $\sigma^{2}$ and
$R\left(\Delta t\right)$ of voltage magnitudes increase as the system
approaches the bifurcation. However, not all of these signs appear
sufficiently early to detect the bifurcation and take mitigating actions.
For example, $\sigma_{\Delta V}^{2}$ in buses 7, 14, and 26 exhibits
a conspicuous increase when the load level is 10--15$\%$ below the
bifurcation. These variables are good early warning signs (EWS) of
the impending bifurcation. In contrast, $\sigma_{\Delta V}^{2}$ in
buses 20 and 36 is not a useful warning sign as its increase occurs
too close to the bifurcation. The situation with autocorrelation is
reversed, as shown in the second panel of Fig.~\ref{fig:arvar_vmag_39bus}.

By examining $\sigma^{2}$ and $R\left(\Delta t\right)$ for all buses
in our test system, we have concluded that, as Fig.~\ref{fig:arvar_vmag_39bus}
illustrates, good EWS occur in two different types of buses. We found
that $\sigma^{2}$ is a good EWS for load buses, whereas $R\left(\Delta t\right)$
is a good EWS at buses that are close to generators with low inertia.
In addition, we found that $\sigma_{\Delta V}^{2}$ at generator buses
is much smaller than at load buses, largely due to generator voltage
control systems. \textcolor{black}{As explained in Sec.~\ref{sec:Detectability},
this limits the use of }$R_{\Delta V}\left(\Delta t\right)$\textcolor{black}{{}
at generator buses as an EWS.}

\subsection{Autocorrelation and Variance of Line Currents \label{sub:Autoco-Var-Currents}}

\begin{figure}[H]
\begin{centering}
\includegraphics[width=1\columnwidth,height=0.2\textheight]{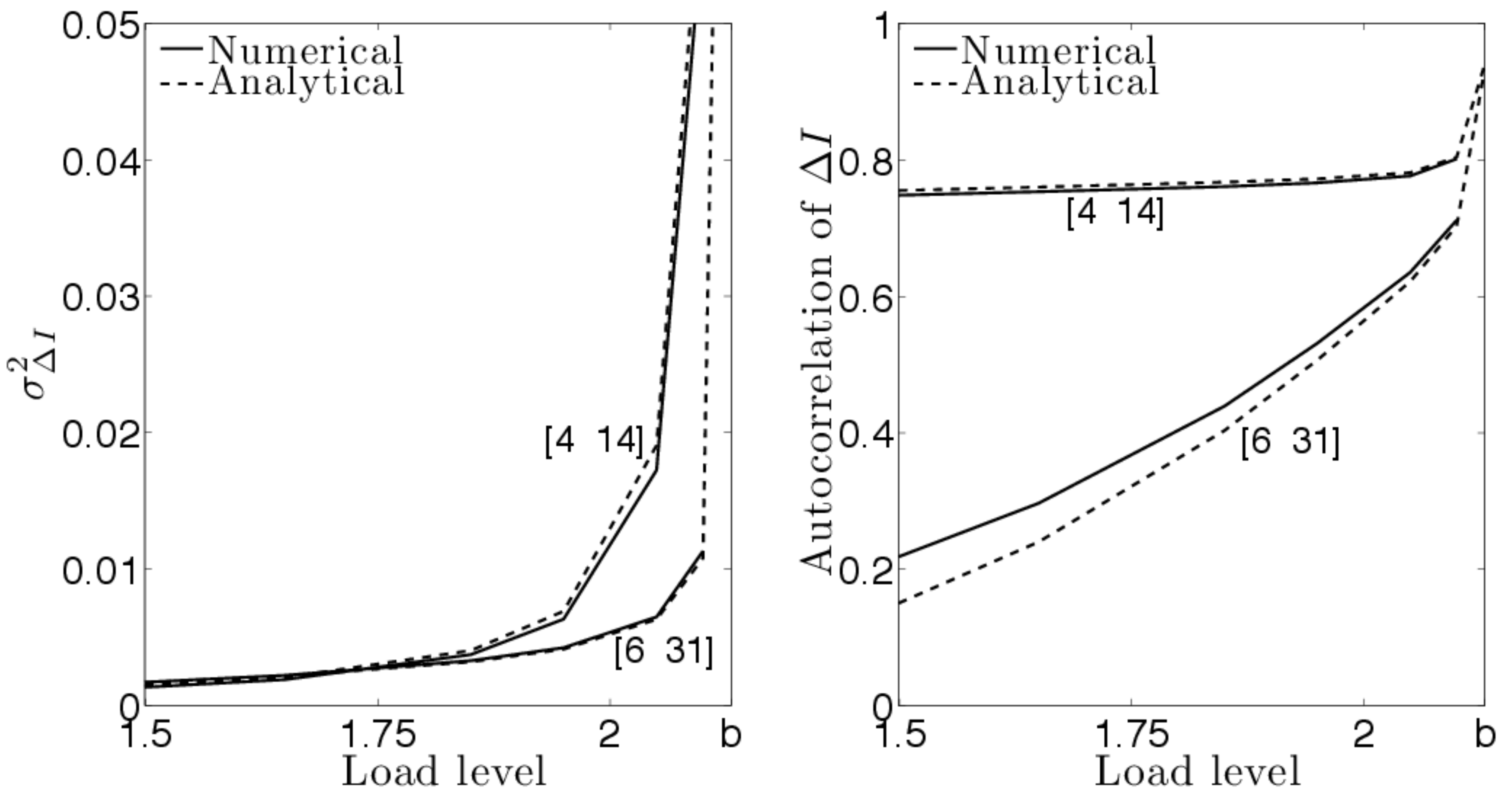}
\par\end{centering}

\vspace{-.1in}

\protect\caption{\label{fig:Varar_Il946}Variance and autocorrelation of current of
two lines. The numbers in brackets are bus numbers at two ends of
the lines. }

\vspace{-.1in}
\end{figure}

The fact that autocorrelation of voltages is not uniformly useful
as an EWS prompted us to look at other variables, particularly currents,
that might be more useful indicators. Results for $\sigma^{2}$ and
$R\left(\Delta t\right)$ of currents, shown in Fig.~\ref{fig:Varar_Il946},
suggest that while $\sigma_{\Delta I}^{2}$ of almost all lines increase
measurably with the increase of the load level, increased $R_{\Delta I}(\Delta t)$
is clearly observable only in some of the lines, such as line $\left[6\,\,31\right]$.
As was the case with voltages, the common characteristic of lines
that show clear increases in $R_{\Delta I}(\Delta t)$ is that they
are connected to a generator with low or moderate inertia. The explanation
for this appears to be that increased $R_{\Delta I}\left(\Delta t\right)$
is closely tied to the way that generators respond to perturbations
as the system approaches bifurcation. Increases in $R_{\Delta I}(\Delta t)$
are not clearly observable in lines that are close to load centers,
such as line $\left[4\,\,14\right]$ in Fig.~\ref{fig:Varar_Il946}.

Examining changes in $\sigma^{2},\, R\left(\Delta t\right)$ of several
state variables showed that only magnitudes of voltages and line currents
signal the proximity to the bifurcation well under certain conditions
mentioned above. Other variables such as voltage angle, current angle,
generator rotor angle and generator speed did not show measurable
or clear monotonically increasing patterns in $\sigma^{2},\, R\left(\Delta t\right)$
that can indicate proximity to a bifurcation.

\section{Detectability after measurement noise\label{sec:Detectability}}

This section examines the detectability of increases in $\sigma^{2}$
and $R\left(\Delta t\right)$ of voltages and currents given the presence
of measurement noise. In addition, we present a method for reducing
the impact of measurement noise using a band-pass filter.

\subsection{Impact of Measurement Noise on Variance and Autocorrelation\label{sub:Results-Meas_Noise}}

Clearly, measurement noise will adversely impact the observability
of increases in $\sigma^{2},\, R\left(\Delta t\right)$ of voltages
and currents. In order to model this impact, we assumed that measurement
noise at each bus is normally distributed with a standard deviation
that is proportional to the steady-state mean voltage for this load
level: $\sigma_{\eta}=0.01\left\langle V\right\rangle $. As a result
the measured variance, $\sigma_{\Delta V_{m}}^{2}$, of a bus voltage
increases to:
\begin{equation}
\sigma_{\Delta V_{m}}^{2}=\sigma_{\Delta V}^{2}+\sigma_{\eta}^{2}\label{eq:meas_volt}
\end{equation}
where $\sigma_{\Delta V}^{2}$ is the variance before adding measurement
noise. 

\begin{figure}[h]
\begin{centering}
\includegraphics[width=1\columnwidth,height=0.22\textheight]{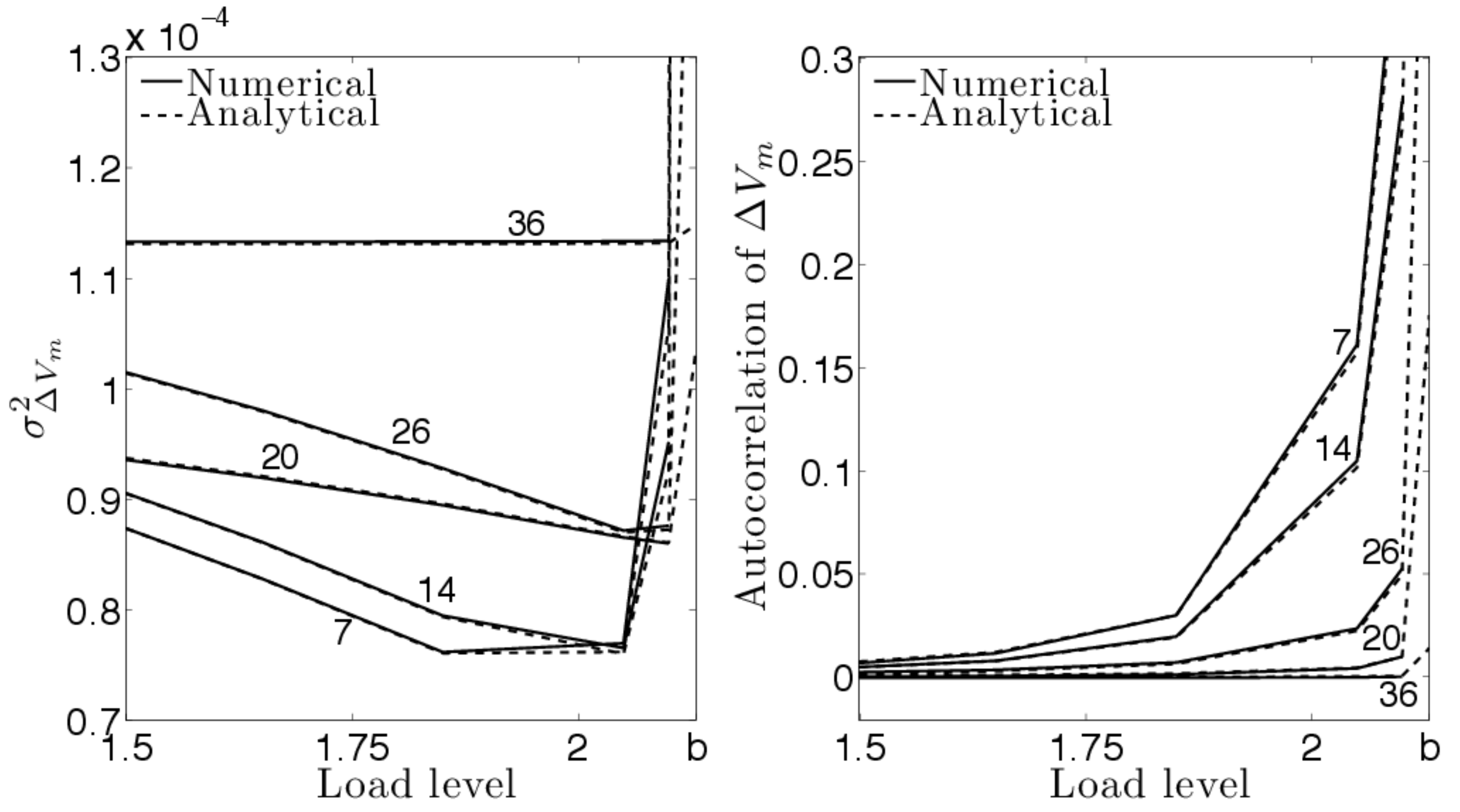}
\par\end{centering}

\vspace{-.1in}

\protect\caption{\label{fig:arvar_Vmag_39bus_nos}Variances and autocorrelations of
voltage magnitudes of five buses in the 39-bus test case versus load
level, accounting for measurement noise.}

\vspace{-.1in}
\end{figure}

Applying this method, Fig.~\ref{fig:arvar_Vmag_39bus_nos} shows
$\sigma^{2}$ and $R\left(\Delta t\right)$ for the voltage magnitudes
of the same five buses used in Sec.~\ref{sub:Autoco-Var-Voltages},
but after adding measurement noise.  The results show that measurement
noise causes the increases in $\sigma_{\Delta V_{m}}^{2}$ to occur
only close to the bifurcation, except for bus 36. In fact, $\sigma_{\Delta V_{m}}^{2}$
decreases for most buses, until close to the bifurcation. \textcolor{black}{The
reason for this decrease is that, based on (\ref{eq:meas_volt}),
$\sigma_{\eta}^{2}$ decreases with $\left\langle V\right\rangle $,
and $\left\langle V\right\rangle $ decreases as the system moves
toward the nose of the PV curve. }Also, because of the 1\% measurement
noise, $\sigma_{\eta}^{2}>\sigma_{\Delta V}^{2}$ until close to the
bifurcation for most buses. For bus 36, which is a generator bus,
$\sigma^{2}$ is almost constant since $\left\langle V\right\rangle $
(and as a result of $\sigma_{\eta}^{2}$) is held constant by the
exciter; $\sigma_{\eta}^{2}\gg\sigma_{\Delta V}^{2}$ for generator
buses. 

Fig.~\ref{fig:arvar_Vmag_39bus_nos} also shows that $R_{\Delta V_{m}}\left(\Delta t\right)$
increases significantly near the bifurcation for buses 7, 14 and,
to a lesser extent, for bus 26. Appendix \ref{meas_nos_artefact}
demonstrates \textcolor{black}{that the increase in }$R_{\Delta V_{m}}\left(\Delta t\right)$\textcolor{black}{{}
of these buses }is largely an artifact of adding measurement noise:
it is primarily due to increases in $\sigma^{2}$ rather than that
of $R\left(\Delta t\right)$. Autocorrelation of $\Delta V_{m}$ is
almost zero for buses 20, 36 since the uncorrelated measurement noise
dominates the voltage of buses near generators.

Thus, measurement noise essentially washes out the useful EWS that
we reported in Sec.~\ref{sub:Autoco-Var-Voltages}. In addition,
there is another issue impacting the detectability of EWS, which we
discuss in the next subsection.

\subsection{Spread of Statistics\label{sub:Spread-of-Statistics}}

One important point regarding the detection of increased $\sigma^{2}$
and $R\left(\Delta t\right)$ is that the measured statistics of a
\textit{sample} of a variable's measurement data (which an operator
can observe in finite time) are different from the mean statistical
properties of that variable over infinitely many measurements. Although
the mean of these statistics typically grows as the system approaches
a bifurcation, the variance (spread) of these statistics that results
from finite sample sizes can cause difficulty in estimating the distance
to the bifurcation. 

\begin{figure}
\begin{centering}
\includegraphics[width=1\columnwidth,height=0.2\textheight]{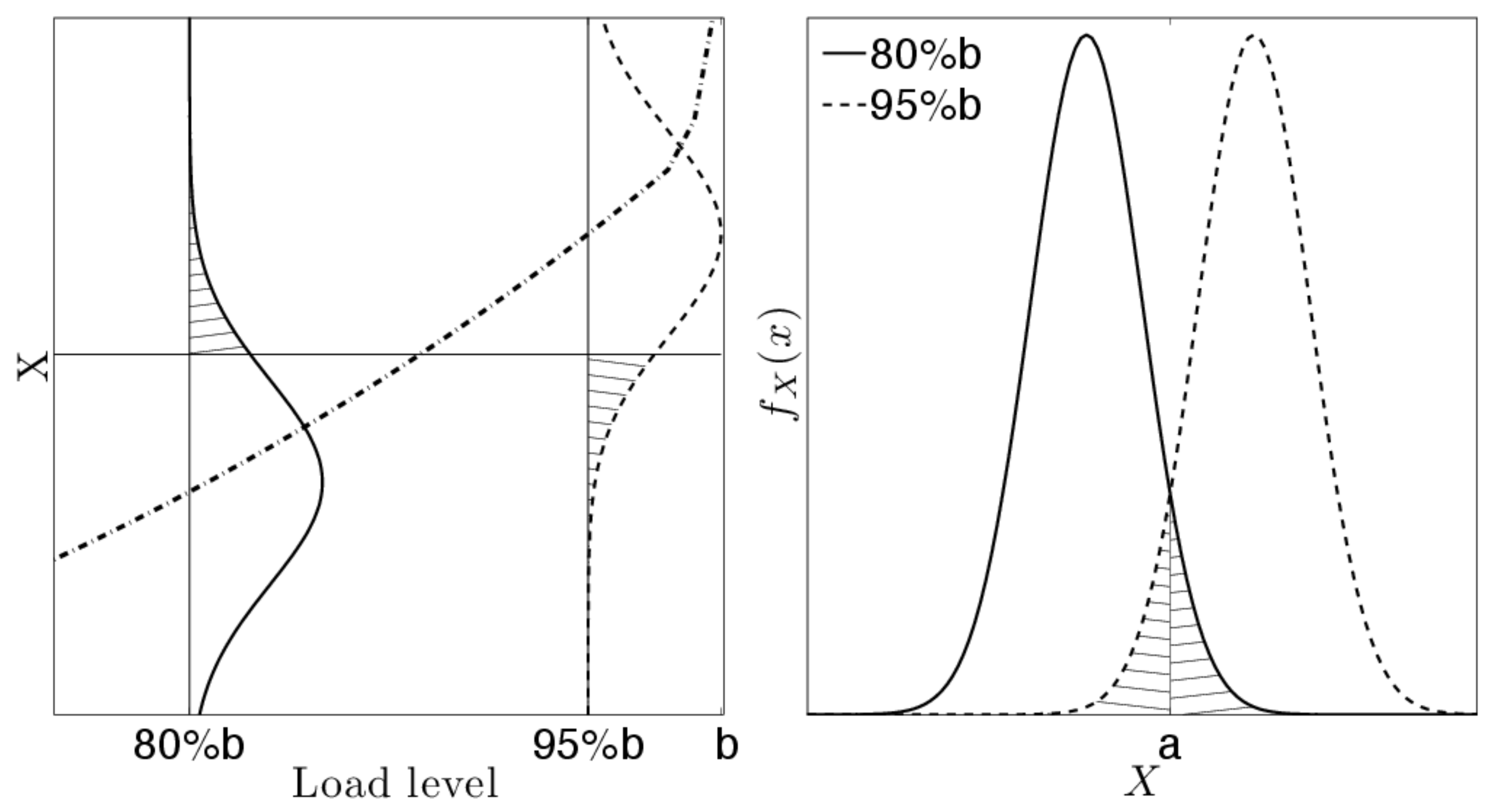}
\par\end{centering}

\vspace{-.1in}

\protect\caption{\label{fig:q9580_idx}The left panel shows the empirical pdfs of $X$,
which can be $\sigma^{2}\,\textnormal{{or}}\, R\left(\Delta t\right)$
of measurements for two load levels. Measure $q_{95/80}$ is equal
to the sum of the hatched areas. The dash-dot line shows the mean
of $X$ versus load level. The right panel shows an alternative view
of the pdfs. }

\vspace{-.1in}
\end{figure}

\textcolor{black}{In order to }quantify the detectability of an increase
in $\sigma^{2}\,\textnormal{{or}}\, R\left(\Delta t\right)$, we introduce
an index $q_{95/80}$ (see Fig.~\ref{fig:q9580_idx}):

\begin{equation}
q_{95/80}=\int_{a}^{\infty}f_{X\left(80\%\right)}dx+\int_{-\infty}^{a}f_{X\left(95\%\right)}dx\label{eq:q9580}
\end{equation}
where $X$ is the statistic of interest ($\sigma^{2}\,\textnormal{{or}}\, R\left(\Delta t\right)$),
$f_{X\left(80\%\right)}\,\textnormal{{and}}\, f_{X\left(95\%\right)}$
are the probability density functions (pdfs) of $X$ for load levels
of $80\%\,\textnormal{{and}}\,95\%$ of the bifurcation, and $a$
is the point where the two distributions intersect. This measure ranges
from 0 to 1, where $0$ suggests that there is no overlap between
the two distributions, such that detectability is unimpeded by the
statistic's spread, while $q_{95/80}=1$ means that the two distributions
completely overlap---i.e. the statistic does not increase. When the
statistic has a decreasing trend, we declare $q_{95/80}=NA$. $q_{95/80}$
roughly corresponds to the probability of being able to correctly
distinguish between the measured statistics at 80\% and 95\% load
levels.

\subsection{Filtering Measurement Noise\label{sub:Band-pass-Filtering} }

\begin{figure}
\begin{centering}
\includegraphics[width=1\columnwidth,height=0.2\textheight]{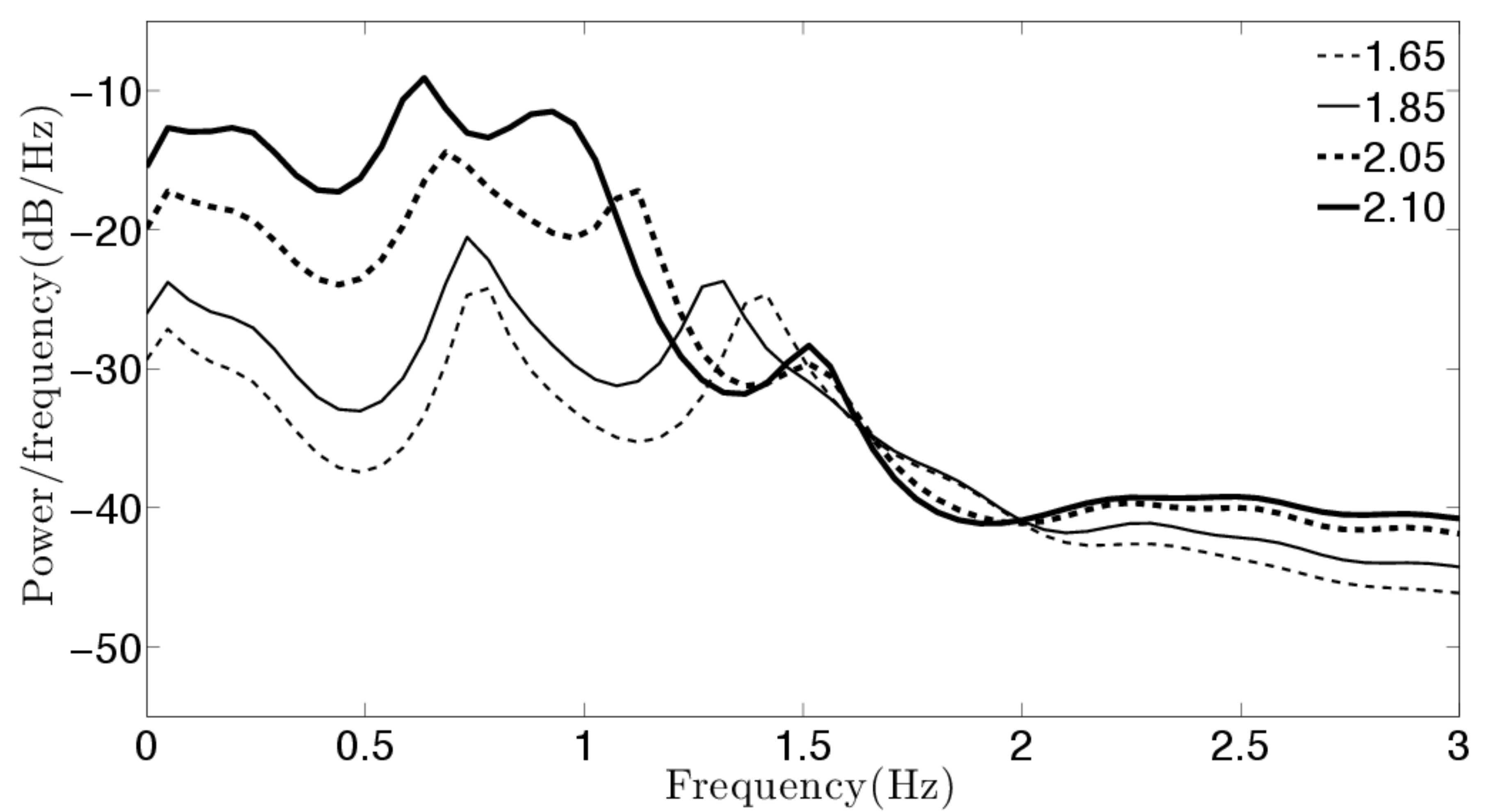}
\par\end{centering}

\vspace{-.1in}

\protect\caption{\label{fig:PSD_Il46}Power spectral density of the current of line
{[}6 31{]} for several load levels listed in the legend. Bifurcation
is at load level=2.12.}

\vspace{-.1in}
\end{figure}

In this section, we explore the use of a band-pass filter to reduce
the impact of \textcolor{black}{{} measurement noise on the statistics
of voltage and current measurements. The reason for filtering out
the high frequency content of measurements is that th}e power spectral
density (PSD) of voltages and currents (see Fig.~\ref{fig:PSD_Il46})
shows that the power of the system noise (i.e., voltage or current
magnitude variations in response to load fluctuations) is concentrated
mostly in its low frequencies. This appears to be typical for Hopf
and saddle-node bifurcations in power systems. On the other hand,
in order to detect CSD, it is necessary to remove slow trends that
result not from CSD but from other factors, such as gradual changes
in the system's operating point \cite{dakos2008slowing}. By experimentation,
we found that a band-pass filter with a pass-band of {[}0.1, 2{]}
Hz reduces the impact of measurement noise in this system optimally.
The rationale for these bounds can be seen from Fig.~\ref{fig:PSD_Il46},
which shows the PSD of a typical current magnitude in our system.
We use this filter for all ``filtered'' results reported subsequently. 

\begin{figure}
\begin{centering}
\includegraphics[width=1\columnwidth,height=0.2\textheight]{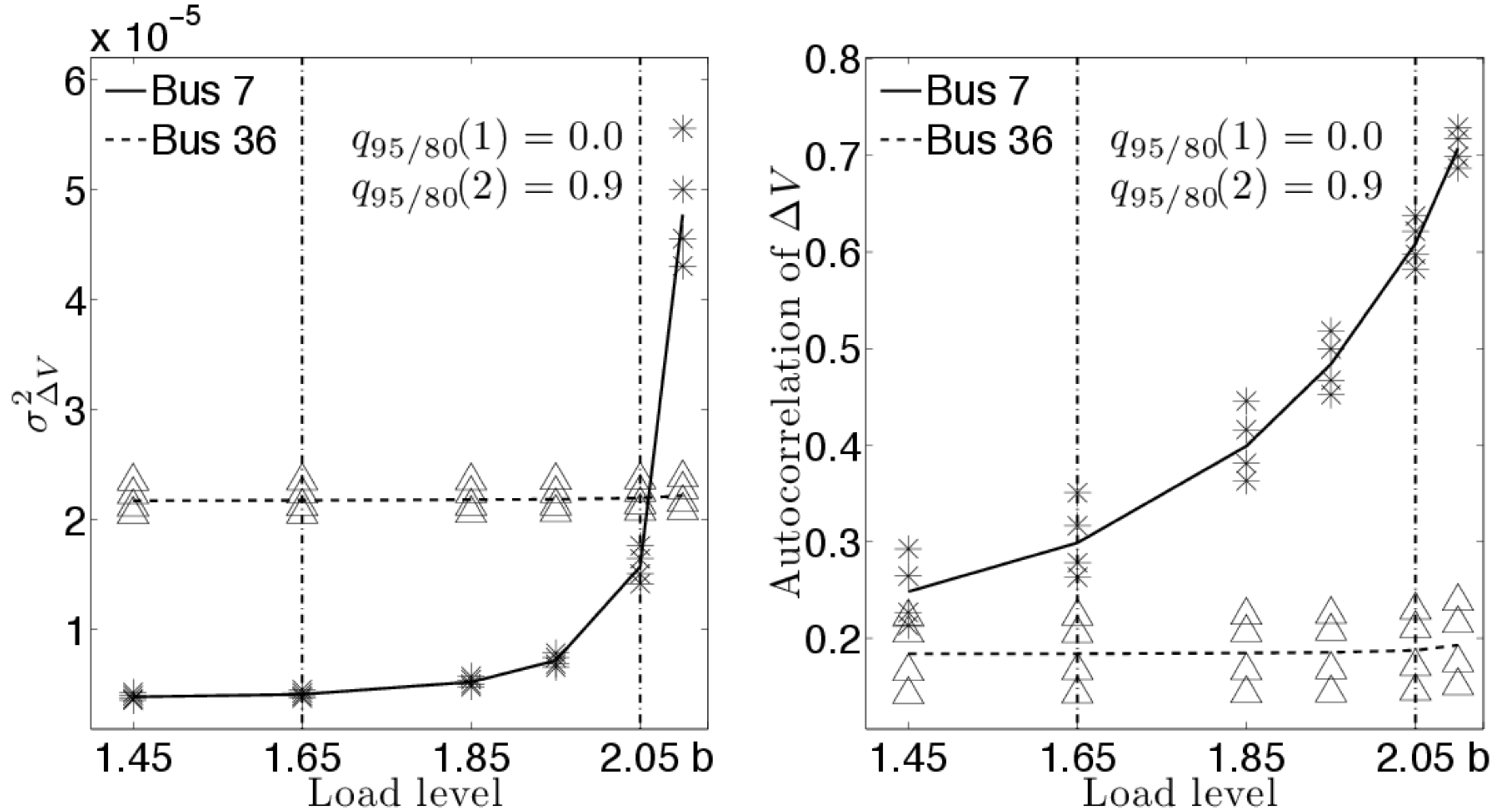}
\par\end{centering}

\vspace{-.1in}

\protect\caption{\label{fig:varar_V736_filt}Variance and autocorrelation of voltage
magnitude of buses 7,36 versus the load level after filtering the
measurement noise. In this and subsequent figures, the lines show
the mean and the discrete symbols $\left(*,\triangle\right)$ represent
5th, 25th, 75th, 95th percentiles of values of $\sigma^{2},\, R\left(\Delta t\right)$
for 100 realizations at each load level. The vertical dash-dot lines
show $Load\, level=80\%b,\,95\%b$.}

\vspace{-.1in}
\end{figure}

Fig.~\ref{fig:varar_V736_filt} shows $\sigma_{\Delta V}^{2},\, R_{\Delta V}\left(\Delta t\right)$
of buses 7, 36 after filtering measurement noise. Comparing Fig.~\ref{fig:varar_V736_filt}
with Fig.~\ref{fig:arvar_Vmag_39bus_nos} shows that using the band-pass
filter significantly improves the detectability of increases in $\sigma_{\Delta V_{7}}^{2}$,
which is close to load centers, but is not effective for bus 36, which
is connected to a generator. The reason is that, even with filtering,
it is still necessary that $\sigma^{2}$ without measurement noise
be sufficiently large so that measurement noise does not dominate
it. $\sigma^{2}$ of measurement noise after filtering will approximately
be:
\begin{equation}
\sigma_{\eta f}^{2}=\sigma_{\eta}^{2}\cdot\nicefrac{\left(f_{H}-f_{L}\right)}{\left(\nicefrac{f_{s}}{2}\right)}\label{eq:var_nos_filt}
\end{equation}
where $\sigma_{\eta f}^{2}$ is the variance of measurement noise
after filtering; $f_{H},f_{L}$ are upper and lower cut-off frequencies
of the filter; and $f_{s}$ is the sampling frequency of measurements.
Assuming $\sigma_{\eta}^{2}=1e-4\,\textnormal{{and}}\, f_{s}=60\textnormal{{Hz}}$,
we get $\sigma_{\eta f}^{2}=6.3\times10^{-6}$. From Fig.~\ref{fig:arvar_vmag_39bus},
one can see that only $\sigma_{\Delta V}^{2}$ of the load buses exceeds
this value near the bifurcation. 

Fig.~\ref{fig:varar_V736_filt} also shows that after filtering out
measurement noise, the increase in $R_{\Delta V_{7}}\left(\Delta t\right)$
is detectable near the bifurcation. However, as mentioned in Sec.~\ref{sub:Results-Meas_Noise},
increases in $R(\Delta t)$ primarily result from increases in $\sigma_{\Delta V}^{2}$,
and thus do not provide additional information regarding the proximity
of the system to the bifurcation. Since $\sigma_{\eta}^{2}\gg\sigma_{\Delta V}^{2}$
for generator buses, $R_{\Delta V_{36}}\left(\Delta t\right)$ also
does not increase measurably as the system approaches the bifurcation,
even after filtering. 

\begin{figure}
\begin{centering}
\includegraphics[width=1\columnwidth,height=0.2\textheight]{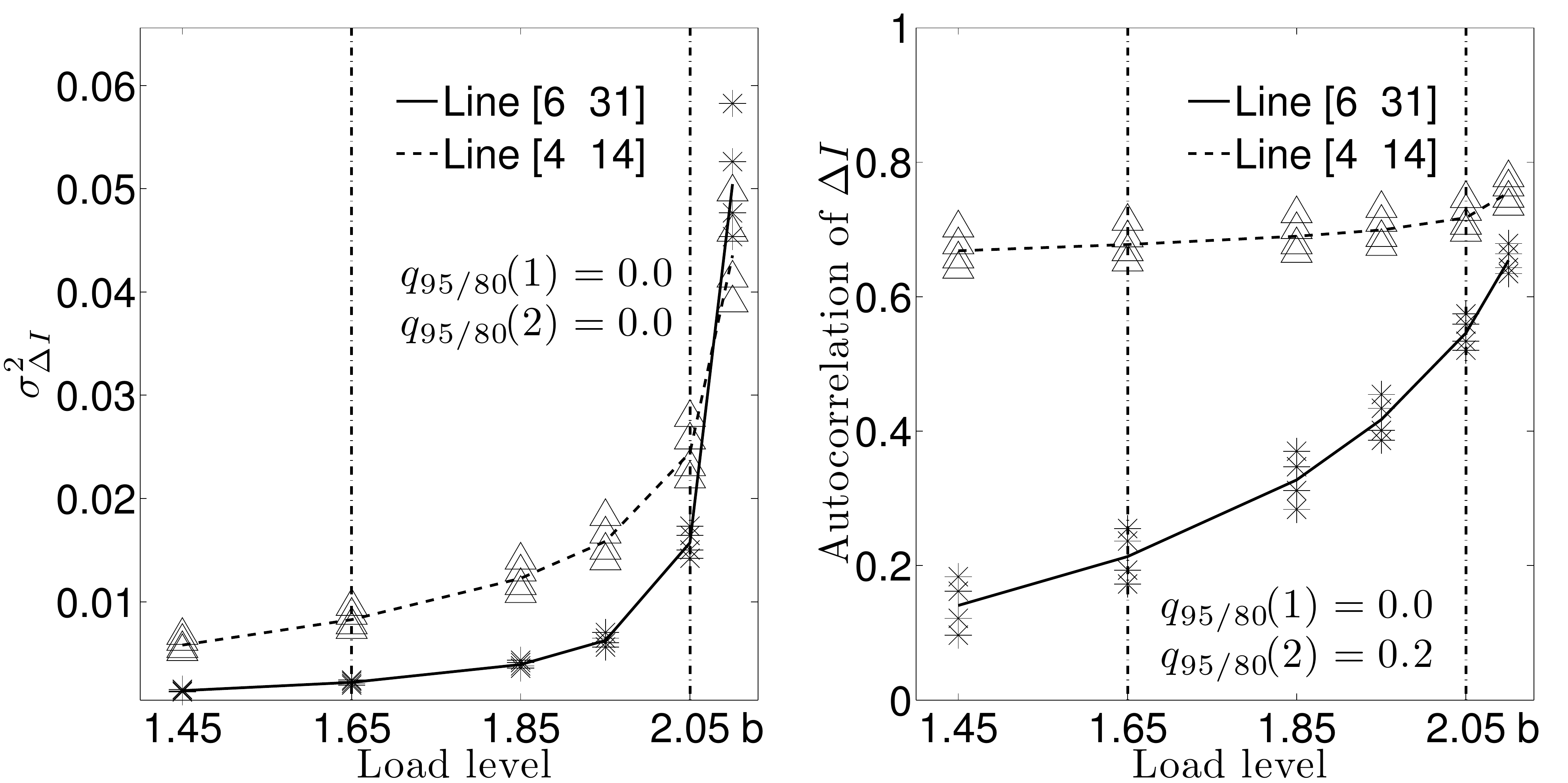}
\par\end{centering}

\vspace{-.1in}

\protect\caption{\label{fig:varar_Il469_filt}Variance and autocorrelation of currents
of lines $\left[6\,\,31\right]$, $\left[4\,\,14\right]$ after filtering
the measurement noise.}

\vspace{-.1in}
\end{figure}

Similar to the case without measurement noise, $R\left(\Delta t\right)$
of line currents close to generators increase more clearly than that
of lines near load centers. Fig.~\ref{fig:varar_Il469_filt} shows
$\sigma^{2},\, R\left(\Delta t\right)$ of currents of lines $\left[6\,\,31\right]$
and $\left[4\,\,14\right]$ after filtering the noise. 

In general, filtering noise from line currents is easier than from
voltages since the ratio of $\sigma^{2}$ of the system noise (defined
above) to $\sigma^{2}$ of measurement noise is larger for currents. 

\begin{figure}
\begin{centering}
\includegraphics[width=1\columnwidth,height=0.2\textheight]{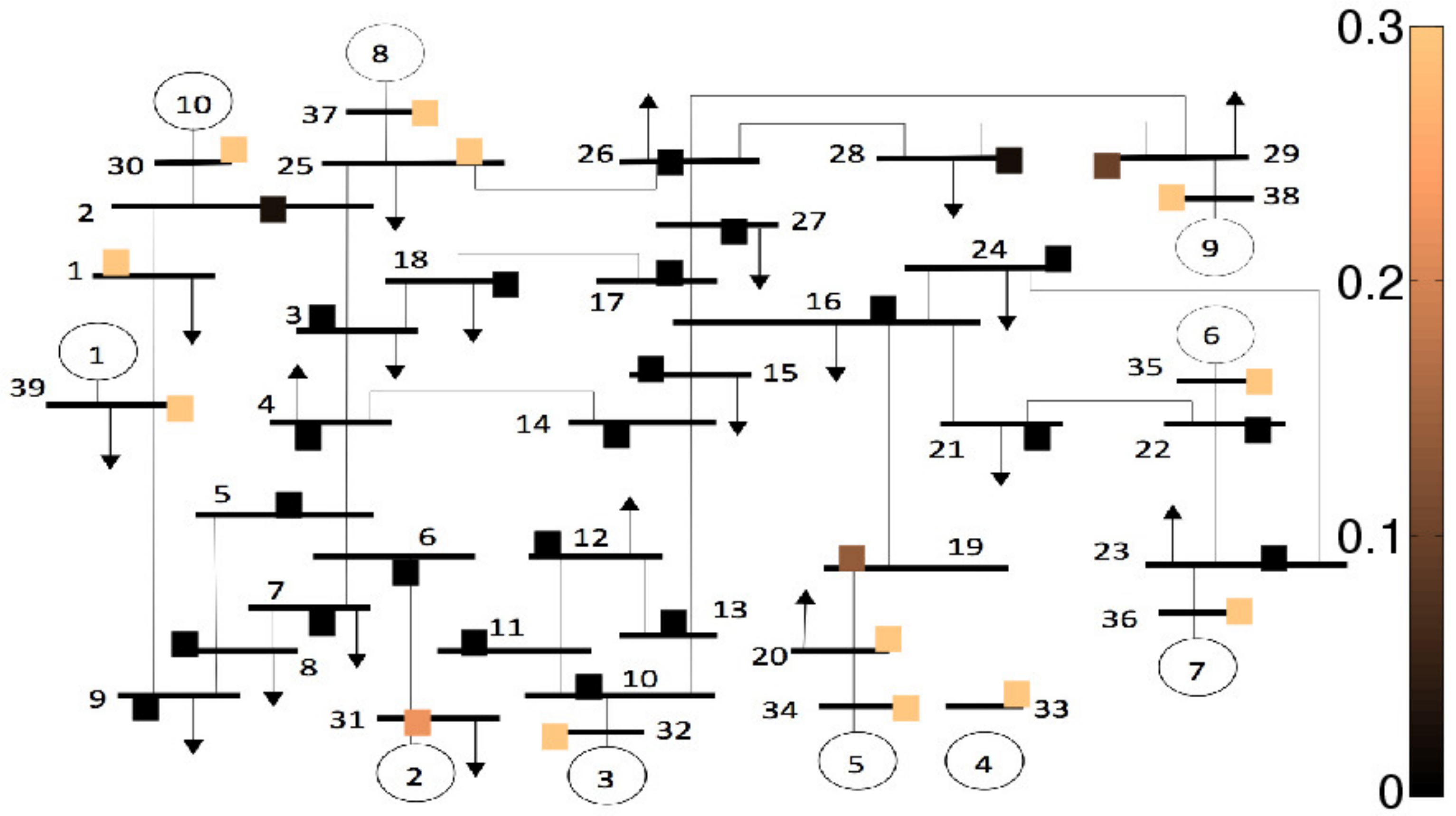}
\par\end{centering}

\vspace{-.1in}

\protect\caption{\label{fig:varv_grid}Index $q_{95/80}$ for $\sigma_{\Delta V}^{2}$
of bus voltages across the 39-bus test case. Here, and in Fig.~\ref{fig:arI_grid},
each rectangle represents the index $q_{95/80}$ for $\sigma_{\Delta V}^{2}$
of the bus next to it. In order to illustrate the results more clearly,
we show $q_{95/80}=0.3$ for measurements with $q_{95/80}>0.3$, because
quantities with this spread become indistinguishable.}

\vspace{-.1in}

\end{figure}

Fig.~\ref{fig:varv_grid} shows the index $q_{95/80}$ for $\sigma_{\Delta V}^{2}$
across the 39-bus test case after filtering measurement noise. The
results in Fig.~\ref{fig:varv_grid} illustrate our earlier statement
that $\sigma_{\Delta V}^{2}$ of buses near load centers are good
EWS of the bifurcation while $\sigma_{\Delta V}^{2}$ of generator
buses are not.

\begin{figure}
\begin{centering}
\includegraphics[width=1\columnwidth,height=0.2\textheight]{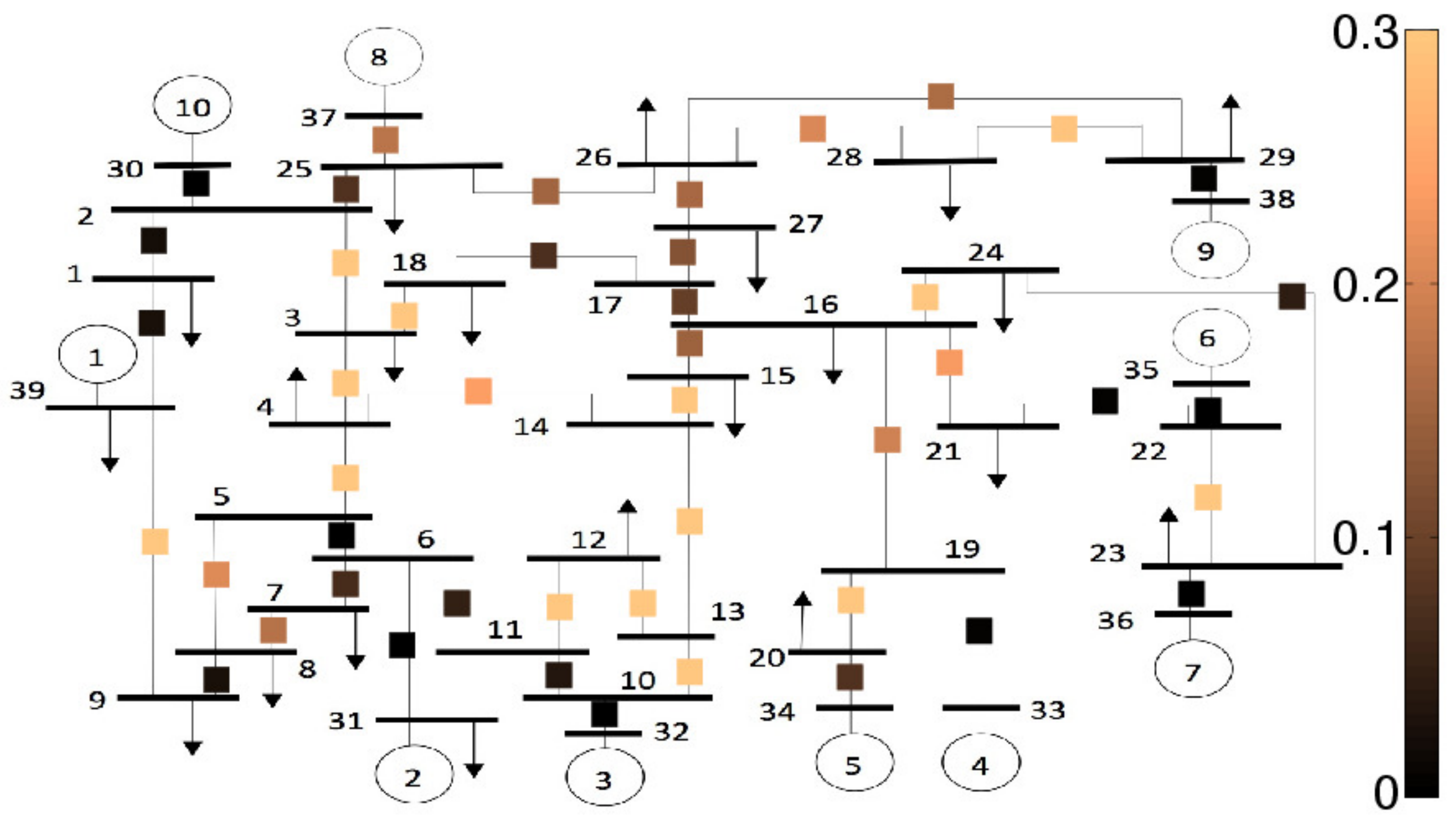}
\par\end{centering}

\vspace{-.1in}

\protect\caption{\label{fig:arI_grid}Index $q_{95/80}$ for $R_{\Delta I}\left(\Delta t\right)$
of lines across the 39-bus test case. Each rectangle represents index
$q_{95/80}$ for $R_{\Delta I}\left(\Delta t\right)$ of the line
next to it.}

\vspace{-.1in}
\end{figure}

Fig.~\ref{fig:arI_grid} shows the index $q_{95/80}$ for $R_{\Delta I}\left(\Delta t\right)$
of lines across the 39-bus test case after filtering the measurement
noise. The results in Fig.~\ref{fig:arI_grid} show that $R_{\Delta I}\left(\Delta t\right)$
of the lines near generators provide good EWS of the bifurcation while
$R_{\Delta I}\left(\Delta t\right)$ of the rest of the lines do not
provide useful EWS.

\textcolor{black}{Note that while filtering of measurement noise can
be helpful in detecting the increase in $\sigma_{\Delta V}^{2}$ of
buses near load centers, it is not helpful in detecting an increase
in $R_{\Delta V}\left(\Delta t\right)$ of these buses. This is because
the $R_{\Delta V}\left(\Delta t\right)$ of such buses are not inherently
good indicators of the proximity to the bifurcation; See Sec.~\ref{sub:Autoco-Var-Voltages}.
Also, filtering measurement noise will not be helpful in retrieving
the statistics of the bus voltages close to generators since their
variances are small compared to that of measurement noise. On the
contrary, $R_{\Delta I}\left(\Delta t\right)$ of lines near generators
provide good EWS for the bifurcation, while $\sigma_{\Delta I}^{2}$
of almost all lines provide good EWS.}

\section{Detecting Locations of Increased Stress\label{sec:Stressed-Area}}

This section examines the potential to use statistical properties
of measurements to detect the locations of increased stress in a power
system. \textcolor{black}{By studying two scenarios, we investigated
whether patterns of change in $\sigma^{2}$ and $R\left(\Delta t\right)$
in a stressed area are different from the rest of the grid, so they
can be helpful in identifying the location of the stressed area.}

\subsection{Transmission line tripping\label{sub:Transmission-line-tripping}}

\begin{figure}
\begin{centering}
\includegraphics[width=1\columnwidth,height=0.2\textheight]{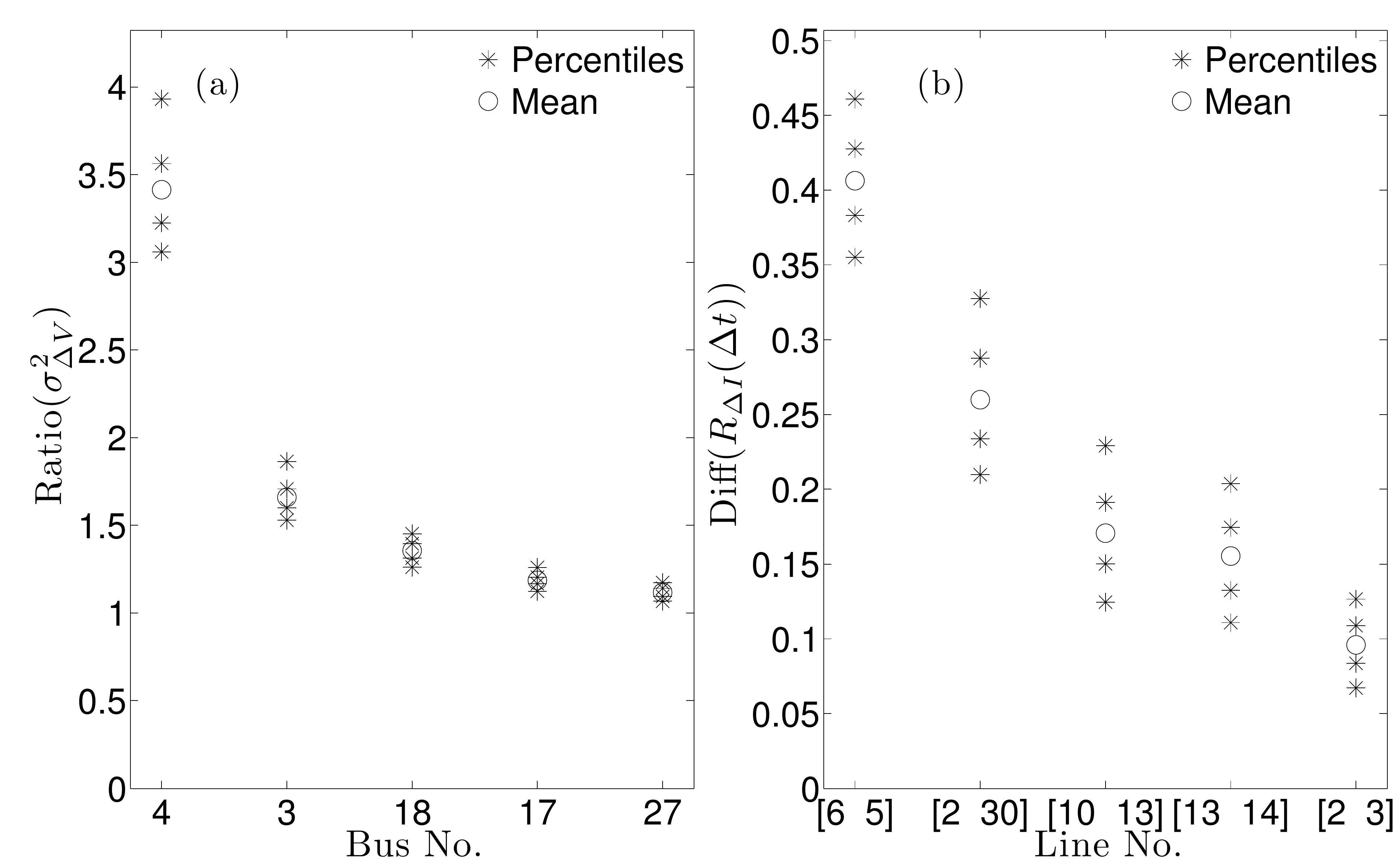}
\par\end{centering}

\vspace{-.1in}

\protect\caption{\label{fig:VarV_Ratio_lines}Panel (a) shows $\textnormal{{Ratio}}\left(\sigma_{\Delta V}^{2}\right)$
after disconnecting the two lines connected to bus 4. The mean of
the $\textnormal{{Ratio}}\left(\sigma_{\Delta V}^{2}\right)$ for
the 5 buses that show the highest increases in variance, as well as
the 5th, 25th, 75th, 95th percentiles of their values, are shown.
Panel (b) shows $\textnormal{{Diff}}\left(R_{\Delta I}\left(\Delta t\right)\right)$
for 5 lines that exhibit the largest increases in $R_{\Delta I}\left(\Delta t\right)$.
The results are shown after filtering of measurement noise.}

\vspace{-.1in}
\end{figure}

In the first scenario, we disconnected lines between buses 4, 14 and
buses 4, 5 in order to increase stress in the area close to bus 4.
For this experiment, the load level was held constant at 1.45 times
the nominal. We calculated the ratio of $\sigma_{\Delta V}^{2}$ and
$\sigma_{\Delta I}^{2}$ for the stressed case to the variances at
the normal operating condition $\left(\textnormal{{Ratio}}\left(\sigma^{2}\right)\right)$.
We also calculated the difference between $R_{\Delta V}\left(\Delta t\right)$
and $R_{\Delta I}\left(\Delta t\right)$ for the two cases $\left(\textnormal{{Diff}}\left(R\left(\Delta t\right)\right)\right)$.
Values of $\textnormal{{Ratio}}\left(\sigma^{2}\right),\,\textnormal{{Diff}}\left(R\left(\Delta t\right)\right)$
that are sufficiently larger than 1 or 0 indicate significant increase
in $\sigma^{2}$ or $R\left(\Delta t\right)$, respectively. Fig.~\ref{fig:VarV_Ratio_lines}(a)
shows $\textnormal{{Ratio}}\left(\sigma_{\Delta V}^{2}\right)$ after
adding measurement noise and filtering. The five bus voltages shown
have the highest mean of $\textnormal{{Ratio}}\left(\sigma^{2}\right)$
among all buses. The figure shows that the voltage of the buses near
bus 4 have the largest $\textnormal{{Ratio}}\left(\sigma_{\Delta V}^{2}\right)$
among the system buses. As with voltages, $\sigma_{\Delta I}^{2}$
close to bus 4 showed more growth than $\sigma_{\Delta I}^{2}$ in
the rest of the system. These results suggest that larger increases
in $\sigma_{\Delta V}^{2}$ and $\sigma_{\Delta I}^{2}$ in one area
of the system, relative to the rest of the system, can indicate that
this area is stressed. 

Our results from Sec.~\ref{sec:Detectability} identified certain
lines whose autocorrelation of currents can be good EWS of bifurcation.
We now comment on what behavior these autocorrelations exhibit in
this experiment. It turns out that not all of these autocorrelations
show a measurable increase; the five lines whose currents' autocorrelations
show the largest increases are shown in Fig.~\ref{fig:VarV_Ratio_lines}(b).
While it is not possible to pinpoint the location of the disturbance
based only on these statistical characteristics, it is possible to
tell, based on the statistics, that the disturbance has occurred in
a certain area of the network. This knowledge would reinforce the
information obtained from monitoring variances of voltages and currents.
As explained in Sec.~\ref{sub:Results-Meas_Noise}, $R_{\Delta V}\left(\Delta t\right)$
does not provide useful information regarding which areas in the grid
are most stressed.

\subsection{Capacitor tripping\label{sub:Capacitor-tripping}}

\textcolor{black}{This section provides an example in which the statistical
measures, }$\sigma^{2}$ and $R\left(\Delta t\right)$, (at least
partially) indicate the location of stress in the network, but the
mean volta\textcolor{black}{ges $\left\langle V\right\rangle $} do
not change enough to be good indicators. This example was designed
to test the hypothesis that $\sigma^{2}$ and $R\left(\Delta t\right)$
can provide information that is not readily available from the mean
values.

\begin{figure}
\begin{centering}
\includegraphics[width=1\columnwidth,height=0.2\textheight]{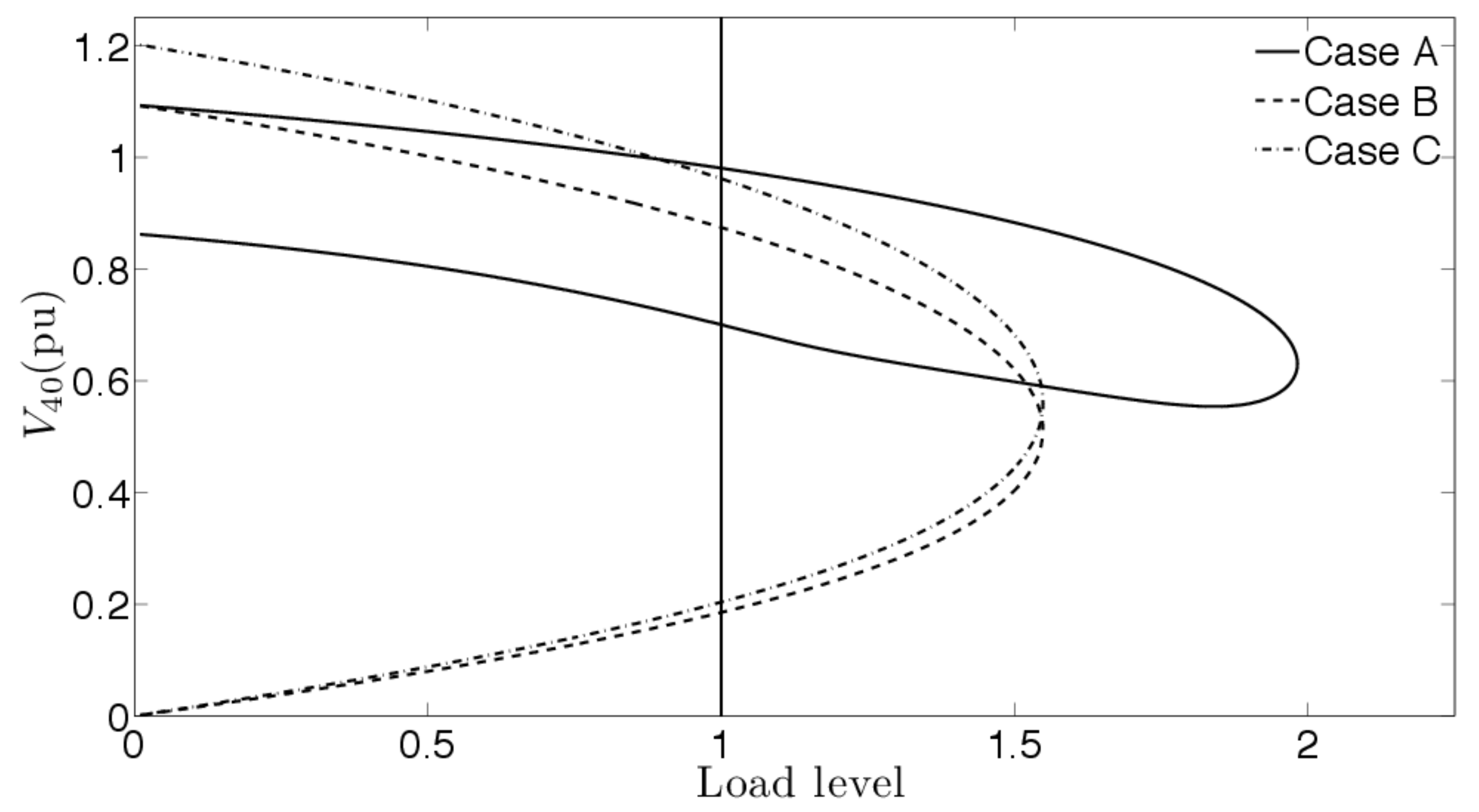}
\par\end{centering}

\vspace{-.1in}

\protect\caption{\label{CPF}PV curve for the three cases described in Sec.~\ref{sub:Capacitor-tripping}.
The vertical line corresponds to the base load level.}

\vspace{-.1in}
\end{figure}

\textcolor{black}{For this example, we added a new bus (bus 40) and
an }under-load tap changing (ULTC) transformer\textcolor{black}{{} that
connects bus 40 with bus 15.} We also transferred the load of bus
15 to bus 40. Fig.~\ref{CPF} shows the P-V curve of bus 40 for three
cases. In Case A, the system is in normal operating condition. In
Case B, a 3-MVAR capacitor at bus 40 is disconnected and in Case C,
the tap changer changes the tap from 1 to 1.1 in order to return the
voltage to the normal operating range ($\left[0.95\,\,1.05\right]$
pu). Fig.~\ref{CPF} shows that the disconnection of the capacitor
reduces the stability margin significantly, which manifests itself
in lower voltage at bus 40. However, the increase in the ULTC's tap
ratio to 1.1 returns the voltage to a value close to its normal level.

\begin{figure}
\begin{centering}
\includegraphics[width=1\columnwidth,height=0.2\textheight]{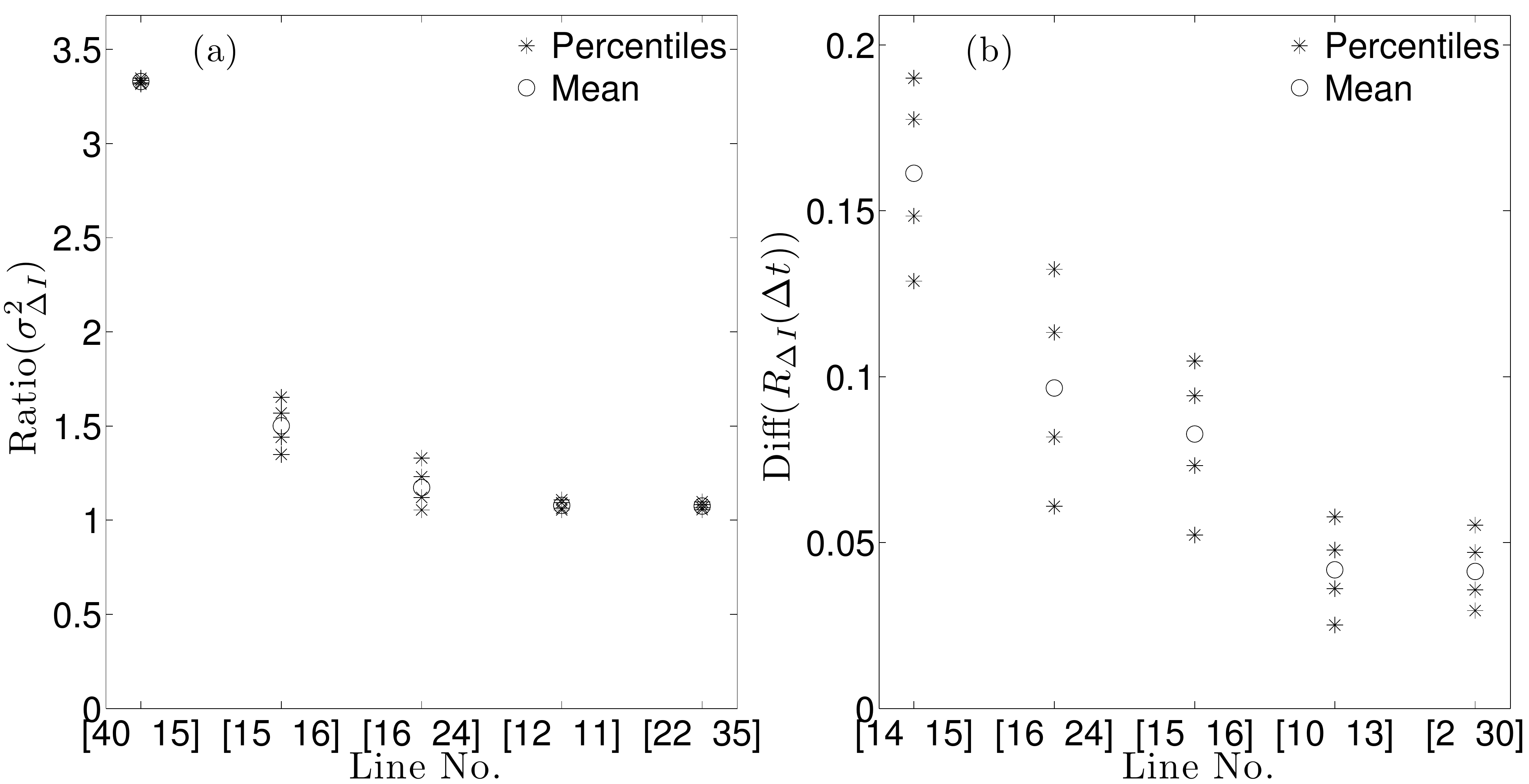}
\par\end{centering}

\vspace{-.1in}

\protect\caption{\label{fig:arvaril_Cap}Panel (a) shows $\nicefrac{{\sigma_{\Delta I,case\, C}^{2}}}{\sigma_{\Delta I,case\, A}^{2}}$
for 5 lines that exhibit the largest increase in $\sigma_{\Delta I}^{2}$
among all lines. Panel (b) shows $R_{\Delta I,,case\, C}\left(\Delta t\right)-R_{\Delta I,,case\, A}\left(\Delta t\right)$
for 5 lines that exhibit the largest increase in $R_{\Delta I}\left(\Delta t\right)$.}

\vspace{-.1in}
\end{figure}

Fig.~\ref{fig:arvaril_Cap}(a) shows $\textnormal{{Ratio}}\left(\sigma_{\Delta I}^{2}\right)=\nicefrac{{\sigma_{\Delta I,case\, C}^{2}}}{\sigma_{\Delta I,case\, A}^{2}}$
for five lines, after filtering the measurement noise. These five
line currents show the largest increase in $\sigma_{\Delta I}^{2}$
among all lines. The first three highest $\textnormal{{Ratio}}\left(\sigma_{\Delta I}^{2}\right)$
occur in lines that are close to the stressed area. However, some
of the lines that are close to that area do not show significant or
any increase in $\sigma_{\Delta I}^{2}$. For example, $\sigma_{\Delta I}^{2}$
of line $\left[\begin{array}{cc}
14 & 15\end{array}\right]$ decreases. Nevertheless, considering lines with the highest growth
in $\sigma_{\Delta I}^{2}$ can clearly be helpful in identifying
the location of the area of the system under excessive stress. As
was the case for line currents, the results show that buses that exhibit
the largest increases in $\sigma_{\Delta V}^{2}$ are close to the
stressed area. Fig.~\ref{fig:arvaril_Cap}(b) shows $\textnormal{{Diff}}\left(R_{\Delta I}\left(\Delta t\right)\right)=R_{\Delta I,,case\, C}\left(\Delta t\right)-R_{\Delta I,,case\, A}\left(\Delta t\right)$
for 5 lines. The positive values indicate the increase in $R_{\Delta I}\left(\Delta t\right)$.
The results in Fig.~\ref{fig:arvaril_Cap}(b) show that lines that
exhibit the largest increase in $R_{\Delta I}\left(\Delta t\right)$
are close to the stressed area.

\subsection{Discussion}

The results presented in this section show that comparing $\sigma_{\Delta V}^{2}$
and $\sigma_{\Delta I}^{2}$ for a stressed operating condition with
their variances for the normal operating condition can be useful in
detecting stressed areas of a power system. The reason for this is
that the variances of voltage and current magnitudes show larger increases
near the stressed area of a power system, compared to variances in
the rest of the system. The results also show that $R_{\Delta I}\left(\Delta t\right)$
can be helpful in detecting the stressed area's approximate location,
although it may not be helpful in pinpointing the exact location of
the stress. Autocorrelation of bus voltages were not found to be useful
for pinpointing the stressed location for the reason explained in
Appendix~\ref{meas_nos_artefact}.

\section{Conclusions\label{sec:Conclusions}}

This paper investigates the use of statistical signals (autocorrelation
and variance) in time-series data, such as what is produced from synchronized
phasor measurement systems, as indicators of stability in a power
system. 

First, we derived a semi-analytical method for quickly computing the
expected autocorrelation and variance for any voltage or current in
a dynamic power system model. Computing the statistics in this way
was shown to be orders of magnitude faster than obtaining the same
result by simulation, and allows one to quickly identify locations
and variables that are reliable indicators of proximity to instability.
Using this method, we showed that the variance of voltage magnitudes
near load centers, the autocorrelation of line currents near generators,
and the variance of almost all line currents increased measurably
as the 39-bus test case approached bifurcation. We found that these
trends persist, even in the presence of measurement noise, provided
that the data are band-pass filtered. Finally, the paper provides
results suggesting that the statistics of voltage and current data
can be helpful in identifying not only whether a system is seeing
increased stress, but also the location of the stress. 

Together, these results suggest that, under certain conditions, these
easily measured statistical quantities in synchrophasor data can be
useful indicators of stability\textit{\textcolor{black}{.}}

\appendices

\section{\label{meas_nos_artefact}}

The equation for $R_{\Delta V_{m}}\left(\Delta t\right)$ before band-pass
filtering is:
\begin{equation}
\nicefrac{\textnormal{E}\left[\Delta V_{m}\left(t\right)\Delta V_{m}\left(s\right)\right]}{\sigma_{\Delta V_{m}}^{2}}=\nicefrac{\textnormal{E}\left[\Delta V\left(t\right)\Delta V\left(s\right)\right]}{\left(\sigma_{\Delta V}^{2}+\sigma_{\eta}^{2}\right)}\label{eq:corr_vm}
\end{equation}
If $\sigma_{\Delta V}^{2}\ll\sigma_{\eta}^{2}$, $R_{\Delta V_{m}}\left(\Delta t\right)$
will be almost zero. This is the case for generator buses or buses
close to generators such as buses $20,36$. However, if $\sigma_{\Delta V}^{2}$
increases such that $\sigma_{\Delta V}^{2}\sim\sigma_{\eta}^{2}$
and $R_{\Delta V}\left(\Delta t\right)$ is sufficiently larger than
$0$ $\left(>0.2\right)$, then $R_{\Delta V_{m}}\left(\Delta t\right)$
will rise significantly with load level, in part because of increase
in $R_{\Delta V}\left(\Delta t\right)$ and in part because of increase
in $\sigma_{\Delta V}^{2}$. This happens for buses close to load
centers such as $7,14$. Comparing $R\left(\Delta t\right)$ of voltage
of buses $7,14$ in Fig.~\ref{fig:arvar_vmag_39bus} with those in
Fig.~\ref{fig:arvar_Vmag_39bus_nos} shows that these quantities
increase significantly after adding measurement noise while their
increase without measurement noise is much smaller. This shows that
the increase in $R_{\Delta V_{m}}\left(\Delta t\right)$ for load
buses is more due to the increase in $\sigma_{\Delta V}^{2}$ than
due to the increase in $R_{\Delta V}\left(\Delta t\right)$.

\bibliographystyle{IEEEtran}
\nocite{*}
\bibliography{Pow_sys}

\section*{Author biographies}

\vspace{-.2in}
\begin{IEEEbiographynophoto}{Goodarz Ghanavati} (S`11) received the B.S. and M.S. degrees in Electrical Engineering from Amirkabir University of Technology, Tehran, Iran in 2005 and 2008, respectively. Currently, he is pursuing the Ph.D. degree in Electrical Engineering at University of Vermont. His research interests include power system dynamics, PMU applications and smart grid.
\end{IEEEbiographynophoto}
\vspace{-.4in}

\begin{IEEEbiographynophoto}{Paul D.~H.~Hines} (S`96,M`07,SM`14) received the Ph.D.~in Engineering and Public Policy from Carnegie Mellon University in 2007 and M.S.~(2001) and B.S.~(1997) degrees in Electrical Engineering from the University of Washington and Seattle Pacific University, respectively.   He is currently an Associate Professor in the School of Engineering, with a secondary appointment in the Dept.~of Computer Science, at the University of Vermont, and a member of the adjunct research faculty of the Carnegie Mellon Electricity Industry Center. Formerly he worked on various electricity industry projects at the U.S.~National Energy Technology Laboratory,  the US Federal Energy Regulatory Commission, Alstom ESCA, and Black and Veatch. He currently serves as the chair of the Green Mountain Section of the IEEE, as the vice-chair of the IEEE PES Working Group on Cascading Failure, and as an Associate Editor for the IEEE Transactions on Smart Grid.  
\end{IEEEbiographynophoto}
\vspace{-.4in}

\begin{IEEEbiographynophoto}{Taras I. Lakoba} received the Diploma in physics from Moscow State University, Moscow, Russia, in 1989, and the Ph.D. degree in applied mathematics from Clarkson University, Potsdam, NY, in 1996. His research interests include the effect of noise and nonlinearity in fiber-optic communication systems, stability of numerical methods, and perturbation techniques.
\end{IEEEbiographynophoto}
\vspace{-.4in}

\vfill
\end{document}